\documentclass[twocolumn,showpacs,preprintnumbers,amsmath,amssymb,aps]{revtex4}

\usepackage{graphicx}
\usepackage{dcolumn}
\usepackage{bm}

\usepackage{color}

\newcommand{\bra}[1]{\langle #1|}
\newcommand{\ket}[1]{|#1\rangle}
\newcommand{\braket}[2]{\langle #1|#2\rangle}

\newcommand{\bfR}{\mathbf{R}}

\newcommand{\bfk}{\mathbf{k}}
\newcommand{\bfq}{\mathbf{q}}

\newcommand{\bfA}{\mathbf{A}}

\newcommand{\XUV}{\text{XUV}}

\newcommand{\IR}{\text{IR}}

\newcommand{\bfeps}{\boldsymbol{\epsilon}}

\begin{document}


\title{Multiple Ionization Bursts in Laser-Driven Hydrogen Molecular Ion}

\author{Norio Takemoto}
\author{Andreas Becker}%
 \affiliation{%
   JILA and Department of Physics, University of Colorado, 440 UCB, Boulder,
   CO 80309-0440
 }%

\date{\today}

\begin{abstract}
Theoretical study on H$_2^+$ in an intense infrared laser field
on the attosecond time-scale reveals that the molecular ion shows multiple
bursts of ionization within a half-cycle of the laser field oscillation,
in contrast to the widely accepted tunnel ionization picture for an atom.
These bursts are found
to be induced by transient localization of the electron at one of the nuclei,
and a relation between the time instants of the localization and
the vector potential of the laser light is derived. Furthermore, an
experimental scheme is proposed to probe the localization dynamics by
an extreme ultraviolet laser pulse.
\end{abstract}

\pacs{32.80.Rm, 33.80.Rv}
\maketitle

\begin{figure}[t]
\begin{center}
\includegraphics[width=.92\linewidth]{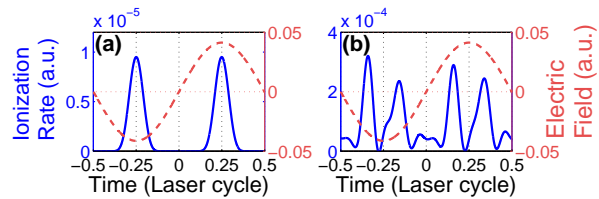}
\caption{
  Time-dependent ionization rates (solid blue curves)
  of (a) H atom \cite{Landau-Lifshitz1977} and (b) H$_2^+$ molecular ion
  over a single period at the peak of a laser pulse, represented by its electric field
  (dashed red curves).
  The ionization rate of H$_2^+$
  is for the range of internuclear distance, $6.75 \text{ a.u.} <R< 7.25
  \text{ a.u.}$, and was obtained by numerical simulation of TDSE.
  Pulse parameters in the calculations were wavelength: $800$ nm,
  peak intensity: $6\times 10^{13}$ W/cm$^2$, pulse duration: $26.69$ fs
  (FWHM).}
\label{fig: tunnel ionization}
\end{center}
\end{figure}

Many interesting and useful phenomena
induced by the intense laser-matter interaction, such as
higher-order harmonics and attosecond pulse generation
as well as non-sequential double ionization,
are known to be initiated by the release of an electron from 
the parent atom or molecule~\cite{prl71_1994}.
This ionization process in an intense laser field is often understood
in terms of the quasi-static tunnel ionization picture.
According to this picture,
the electron tunnels through the barrier created by the combination of the
binding potential of the ionic core and the electric potential of the
laser field.
Thus, in the oscillating electric field of laser light, the electron is expected
to escape most likely at the peak of the electric field strength in every half-cycle
when the barrier becomes the thinnest.
This expectation is in agreement with the temporal dependence of
the ionization rate for an atom 
predicted by the well-known tunneling formula~\cite{Landau-Lifshitz1977,physRevA64_013409}
and shown in Fig.~\ref{fig: tunnel ionization}(a). 
In a recent experiment,
the ionization yield of Ne$^+$ was measured as a function of time 
with attosecond resolution,
and
the results also support the picture above~\cite{nature446_627}.
However,
results of our present theoretical study indicate that
this popular picture needs to be modified for molecules:
they can exhibit not only a single but
multiple bursts of ionization within a half-cycle of
the laser field.
For example, the ionization rate of H$_2^+$
shown in Fig.~\ref{fig: tunnel ionization}(b)
has two maxima within a half-cycle
and minima near the peaks of the laser electric field.

In this Letter we show that these multiple ionization bursts (MIBs) are related
to transient electron localization at one of the protons on the attosecond
time scale.
The localization is due to the strong coupling between a
pair of states with opposite parities,
called the charge-resonant (CR) states~\cite{jcp7_20},
which leads to more complex electron dynamics in a molecule
than in an atom, as it has been reported before \cite{prl101_213002,jcp110_11152}.
Despite the complexity of the electron dynamics, we derive a simple relation
of the instants of electron localization to the vector potential of the
laser field and the coupling strength between the CR states. This enables us
to generalize our observations for the simplest molecule and to predict a
similar modification of the tunnel ionization picture for other molecules
having the CR states as well.
To probe the transient electron localization in
experiments, we propose a method to retrieve the laser-dressed
quantum state of H$_2^+$ from an interference
pattern in the photoelectron momentum distribution generated by applying
an additional attosecond extreme ultraviolet (XUV) pulse.


For the present analysis we have solved the time-dependent Schr\"odinger equation
(TDSE)
for a model of H$_2^+$ in which the
electronic and nuclear motions are restricted along the polarization direction
of the linearly polarized laser pulse.
The Hamiltonian of this system is given by
(Hartree atomic units are used throughout)
\cite{physRev122_1207}:
\begin{equation}\label{hamiltonian}
\begin{split}
&H(t) =
-\frac{1}{2\mu_R}\frac{\partial^2}{\partial R^2}
-\frac{1}{2\mu_z}\frac{\partial^2}{\partial z^2}
  -\frac{1}{\sqrt{\left(z-\frac{R}{2}\right)^2 + a_{\text{e}}}}
\\&
  -\frac{1}{\sqrt{\left(z+\frac{R}{2}\right)^2 + a_{\text{e}}}}
  +\frac{1}{\sqrt{R^2 + a_{\text{n}}}}
+\beta z E(t),
\end{split}
\end{equation}
where $R$ is the internuclear distance, $z$ is the position of the
electron measured from the center-of-mass of the protons,
$\mu_R=M/2$ and $\mu_z=2M/(2M+1)$ are the reduced masses
with $M= 1.836\times 10^{3}$ a.u.\ being the proton mass,
$a_{\text{e}}=1.0$ a.u.\ and $a_{\text{n}}=0.03$ a.u.\ are
the soft-core parameters~\cite{physRevA53_2562},
$\beta=1 + 1/(2M+1)$,
and $E(t)$ is the laser electric field.
It has been shown before that this type of reduced-dimensional models
reproduce experimental results at least qualitatively 
\cite{physRevA53_2562,physRevA65_013402,jPhysB36_3325}.
The ionization rate in Fig.~\ref{fig: tunnel ionization}(b) was
calculated as the out-going probability flux
from the rectangular domain,
$-8.4  \text{ a.u.} < z < 8.4 \text{ a.u.}$ and
$6.75 \text{ a.u.} < R < 7.25 \text{ a.u.}$,
toward $|z|\rightarrow\infty$
normalized by the probability inside this domain,
under 
a laser pulse with the wavelength of 800 nm, peak intensity $6\times 10^{13}$
W/cm$^2$, and FWHM duration $26.69$ fs.

\begin{figure}[t]
\begin{center}
\includegraphics[
        width=0.90\linewidth]
		{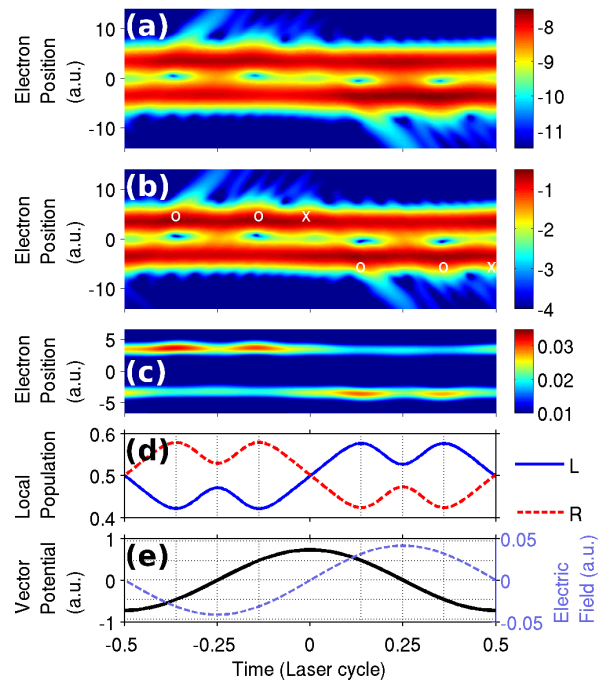}

\caption{
 Electron density 
(a) integrated over $6.75$ a.u.\ $<R<7.25$ a.u.\ in the moving-nuclei model and 
(b) in the fixed-nuclei model at $R=7$ a.u.\ as a function of time and electron
position (shown on a $\log_{10}$-scale).
(c) Electron density in the fixed-nuclei model with absorber near the nuclei.
(d) Electron populations in the localized states, $\ket{\text{L}}$ and
$\ket{\text{R}}$, in the two-state model.
(e) Vector potential (solid line) and electric field (dashed line) of the
laser light. Horizontal and vertical grid lines represent 
$A(t_{\text{loc}})$ and $t_{\text{loc}}$, respectively.
}
\label{fig: Pzt fixed nuc and 2lev}
\end{center}
\end{figure}

We present in Fig.~\ref{fig: Pzt fixed nuc and 2lev}(a) how the electron
density evolves in time under the same laser pulse 
and in the same scope of $R$
as used in Fig.~\ref{fig: tunnel ionization}(b).
This result shows that the electron density bound to the protons
at around $z \approx \pm 3.5$ a.u.
is released in several bunches within a half-cycle of the laser field,
consistently with the observation in Fig.~\ref{fig: tunnel ionization}(b).
In the following, we trace the origin of these MIBs
in the time-evolution of the electron density by simplifying the model.

Figure~\ref{fig: Pzt fixed nuc and 2lev}(b) shows the
electron density calculated by fixing the nuclear positions at $R=7$ a.u.\
and letting $M\rightarrow \infty$. This fixed-nuclei model reproduces
the result of the moving-nuclei model in Fig.~\ref{fig: Pzt fixed nuc and 2lev}(a)
almost perfectly, indicating that the coupling between the electronic and
nuclear motions is not essential for the formation of the MIBs.
In Fig.~\ref{fig: Pzt fixed nuc and 2lev}(c), in the fixed-nuclei
model, we absorbed the ionizing
wavepackets soon after they left the protons by using a $\cos^{1/2}$-mask
function set over $6.6\text{ a.u.} < |z|<11 \text{ a.u.}$
This result shows that the bound electron density is transiently
localized at one of the two protons at the same instants
as the ionization bursts marked by the circles in panel (b).
Part of the wavepackets released in the 
bursts during the time period in which $|E(t)|$ is decreasing
($-0.25<t<0$ and $0.25<t<0.5$ laser cycles 
in Fig.~\ref{fig: Pzt fixed nuc and 2lev})
can return to the protons~\cite{prl71_1994,physRevA49_2117}.
Such rescattering wavepackets 
create the modulation of the density including the enhancements at the cross
marks in Fig.~\ref{fig: Pzt fixed nuc and 2lev}(b) due to the
interference with the bound wavepacket and also  
the finer fringes in the ionizing density at $|z| \gtrapprox 8$ a.u. 
due to the interference with the wavepackets just released from the protons.


From the analysis so far, we can conclude that the MIBs are 
induced by the transient
electron localization at one of the protons while
the (rescattering) dynamics of the electronic wavepackets in the
continuum does not affect the time instants of the MIBs.
This also substantiates that the present 1D models are
sufficient for the analysis since effects of spreading 
in the transverse direction are negligible
for the bound wavepackets.
In previous theoretical results,
the laser-driven electron dynamics \textbf{\it inside} the molecule has been
recognized to be complex and sometimes counter-intuitive~\cite{prl101_213002},
and, in particular, the existence of the sub-laser-cycle electron localization
has been pointed out~\cite{jcp110_11152}.
The present work elucidates that this ultrafast electron localization
also makes the ionization dynamics of H$_2^+$ qualitatively different from that of atoms 
and may give rise to a paradigm change 
from the widely accepted tunneling ionization picture.

In order to further analyze the phenomenon of MIBs, we investigate
the bound state dynamics using a simple model which
incorporates only   
the ground ($\ket{\text{g}}$) and first-excited  ($\ket{\text{u}}$)
electronic states of H$_2^+$.  
These two states 
can be 
written 
as superpositions,
$\ket{\text{g}}=[\ket{\text{L}}+\ket{\text{R}}]/\sqrt{2}$ 
and 
$\ket{\text{u}}=[\ket{\text{L}}-\ket{\text{R}}]/\sqrt{2}$,
of the ground states of H atom, $\ket{\text{L}}$ and $\ket{\text{R}}$, 
centered at the protons at $z=-R/2$ and $z=R/2$, respectively.
During the interaction with laser light,
the eigenstates $\ket{\text{g}}$ and $\ket{\text{u}}$ mix, 
and the electron is driven between the two protons. 
This generates a large transition dipole
$d_{\text{gu}}=-\bra{\text{g}}z\ket{\text{u}}$, which is
proportional to $R$ at large $R$, and is known as the CR mechanism
~\cite{jcp7_20,astrophysJ123_353,physRevA53_2562}.
The two-state model is applicable to the present case since
at $R=7$ a.u.
the $\ket{\text{g}}$ and $\ket{\text{u}}$ states are almost degenerate 
and well isolated from the higher-lying 
excited electronic states on the scale of the IR photon energy. 
The time-evolution of the state $\ket{\Psi(t)}$ of this model in a laser field
can be described in the basis of Floquet states
~\cite{prl67_516,laserPhys3_375,jPhysB34_2371}.
Using a series expansion for the Floquet states~\cite{jPhysB34_2371},
we obtained expressions for the local populations at the respective protons,
$|\braket{\text{L}}{\Psi(t)}|^2$ and $|\braket{\text{R}}{\Psi(t)}|^2$.
From the condition $d[|\braket{\text{L}}{\Psi(t)}|^2]/dt=0$,
we derive that population at one of the protons is maximized at 
the instants $t_{\text{loc}}$ at which the vector potential satisfies
\begin{equation}\label{eq: A tloc}
A(t_{\text{loc}}) =
A_0\sin(\omega t_{\text{loc}}+\varphi) = \frac{m\pi + \chi}{2 d_{\text{gu}}},
\end{equation}
where 
$m=0, \pm 1, \pm 2,\dots$, and 
$\chi$ is a mixing angle determined by the
quantum amplitudes $c_1$ and $c_2$ of the Floquet states 
which reduces to $\ket{\text{g}}$ and $\ket{\text{u}}$ at zero laser intensity,
respectively, with $\cos\chi = (|c_1|^2 - |c_2|^2)/C$,
$\sin\chi = 2\text{Im}[c_1^* c_2]/C$, and
$C=\sqrt{(|c_1|^2 - |c_2|^2)^2 +4(\text{Im}[c_1^* c_2])^2}$.
Furthermore, from the sign of
$d^2|\braket{\text{L}}{\Psi(t_{\text{loc}})}|/dt^2$,
we can predict that the electron density
is localized at the proton down the slope of the electric potential of the
laser field at those $t_{\text{loc}}$ corresponding to odd $m$.
Previously, the electron localization was analyzed
in terms of the phase-adiabatic states,
but numerical computation was necessary to predict the instants
of the maximum localization~\cite{jcp110_11152}.

In Fig.~\ref{fig: Pzt fixed nuc and 2lev}(d), the local populations,
$|\braket{\text{L}}{\Psi(t)}|^2$ and $|\braket{\text{R}}{\Psi(t)}|^2$,
calculated by the two-state model are plotted as a function of time.
Comparison with
Fig.~\ref{fig: Pzt fixed nuc and 2lev}(c)
shows that the two-state model does reproduce the bound
electron dynamics obtained by the numerical
simulation for the 1D fixed-nuclei model.
Therefore, the number and time instants of the sub-cycle ionization bursts
can be traced back to the simple condition (\ref{eq: A tloc}).

We may emphasize that the \textit{strong} and
\textit{exclusive} coupling between a pair of states, and hence the trapping
of the population within such a pair, are the essence of the 
counter-intuitive two-state dynamics inducing the MIBs. 
Thus, the current analysis applies in general to molecules in which 
the ground state is strongly coupled to one particular excited state,
and to laser wavelengths which do not coincide with a resonant transition
to any other state. 
In atoms, the first-excited 
state is coupled not only to the ground state but also with higher-lying 
excited states. Therefore, the electron localization can not be induced, 
and MIBs have not been observed.
Even in H$_2^+$, at around the equilibrium internuclear distance
($R_{\text{eq}}\approx 2$ a.u.), the energy levels of $\ket{\text{g}}$ and
$\ket{\text{u}}$
are not well isolated from the other states, and the transition dipole 
is not exclusive between these two 
states~\cite{jcp7_20,astrophysJ123_353,physRevA53_2562}.
Therefore, the ionization dynamics becomes similar to atoms, and the standard
quasi-static tunnel ionization picture should be recovered.
On the other hand, in H$_2^+$ at larger $R$, where the coupling strength
$d_{\text{gu}}\sim R/2$ becomes large, the electron density can be localized more
than twice within a half-cycle at rather moderate laser intensity.
Note that 
the coupling between CR states can be large and exclusive
already at the equilibrium structure
in other molecules~\cite{pnas25_577}.

As shown above, MIBs are caused by the 
localization of the electron at one of the protons. 
Previously, theoretical studies have shown that
the time-evolving asymmetry of the electron density inside H$_2^+$ can be probed by 
an attosecond XUV pulse via
the asymmetry of the photoelectron yield and momentum 
in opposite directions along the XUV laser polarization
parallel to the molecular axis~\cite{physRevA72_051401,prl101_103001}.
We propose here to extend this scheme by setting
the polarization of the XUV pulse at an angle to the molecular axis and the
polarization of the IR driving pulse and by analyzing the 2D
photoelectron momentum distribution. As we will show below, this potentially enables us to
\textit{reconstruct} the temporal evolution of the \textit{amplitudes} $c_{\text{L}}(t)$ and
$c_{\text{R}}(t)$ of the two quantum states $\ket{\text{L}}$ and
$\ket{\text{R}}$ composing the laser-dressed state of H$_2^+$.

To this end, we model the photoionization process due to the XUV pulse
as a one-photon transition from the IR-laser-dressed bound state,
$\ket{\Psi(t)}
= c_{\text{L}}(t)\ket{\text{L}} + c_{\text{R}}(t) \ket{\text{R}}$,
to the Volkov state of drift momentum $\bfk$ in the IR laser field.
For simplicity, we assume here that the XUV pulse is polarized
perpendicular to the molecular axis and
its electric field has a Gaussian envelope as
${\bf E}_{\XUV}(t)
= \hat{\bfeps}_{\XUV} E^0_{\XUV} \exp\left[ -(t-t_0)^2 /\tau_{\XUV}^2 \right]
\cos(\Omega t + \varphi_{\XUV})$.
At the limit of $\tau_{\XUV} \ll T_{\IR}$, where $\tau_{\XUV}$ is the pulse
duration of the XUV pulse and $T_{\IR}$ is the period of the IR driving pulse,
by applying Laplace's method of asymptotic analysis for the $S$-matrix element,
we obtained the photoelectron momentum distribution as
\begin{equation}
\label{eq: photoelectron distribution by cL and cR}
\begin{split}
&|S_{fi}(\bfk,t_0)|^2
\sim \pi^3 [E^0_{\XUV}]^2 \tau_{\XUV}^2
\left| \tilde{d}_{\text{atom}}(\bfq(t_0)) \right|^2
\\
&\times \exp\left\{ -\frac{\tau_{\XUV}^2}{2} \left[ \frac{|\bfq(t_0)|^2}{2}
	   -\frac{E_{\text{g}}+E_{\text{u}}}{2} - \Omega \right]^2
  \right\}
\\
&\times 
\big\{ |c_{\text{L}}(t_0)|^2 + |c_{\text{R}}(t_0)|^2
\big.\\&\phantom{nnn}
        +2\text{Re}\left[c_{\text{L}}(t_0) c_{\text{R}}^*(t_0) \right]
          \cos[\bfq(t_0)\cdot\bfR]
\\&\phantom{nnn}\big.
        -2\text{Im}\left[c_{\text{L}}(t_0) c_{\text{R}}^*(t_0) \right]
          \sin[\bfq(t_0)\cdot \bfR] \big\}
          ,
\end{split}
\end{equation}
where $\bfq(t_0)=\bfk+\bfA(t_0)$ is the photoelectron momentum $\bfk$ offset by the 
vector potential $\bfA(t_0)$ of the IR field at the peak $t_0$ of the XUV
pulse, $E_{\text{g}}$ and $E_{\text{u}}$ are the energies of
$\ket{\text{g}}$ and $\ket{\text{u}}$, respectively, and 
$ \tilde{d}_{\text{atom}}(\mathbf{q})
=\bra{e^{i\mathbf{q}\cdot \mathbf{r}}}\mathbf{r}\ket{\phi_{\text{1s}}(\mathbf{r})}$ is the
transition dipole between H atom 1s state to the plane wave state.

By fitting 
the model eq.~(\ref{eq: photoelectron distribution by cL and cR})
to an experimentally obtained photoelectron momentum distribution, the quantum amplitudes
$\{c_{\text{L}}(t_0), c_{\text{R}}(t_0)\}$ at the peak of the XUV probe pulse can be retrieved.
In order to demonstrate the accuracy of this retrieval procedure,
we simulated momentum distributions by solving the TDSE
for a 2D model of H$_2^+$ with the internuclear distance fixed at $R=7$ a.u.
To fulfill the condition $\tau_{\XUV} \ll T_{\IR}$, we used an
IR laser pulse with wavelength of $1400$ nm, 
peak intensity $1.5\times 10^{13}$ W/cm$^2$, and FWHM pulse duration
$14.01$ fs. For the XUV laser field, the peak intensity was set at
$1.0\times 10^{12}$ W/cm$^2$, wavelength at $25$ nm, and FWHM
at $500.3$ as.
By changing the delay $\Delta t$ from the peak of the IR pulse to that of XUV
pulse, photoelectron momentum distributions were calculated.
Then, from these distributions,
the background momentum distribution calculated by applying only the IR laser pulse
was subtracted.

\begin{figure}[t]
\begin{center}
\includegraphics[width=.95\linewidth]{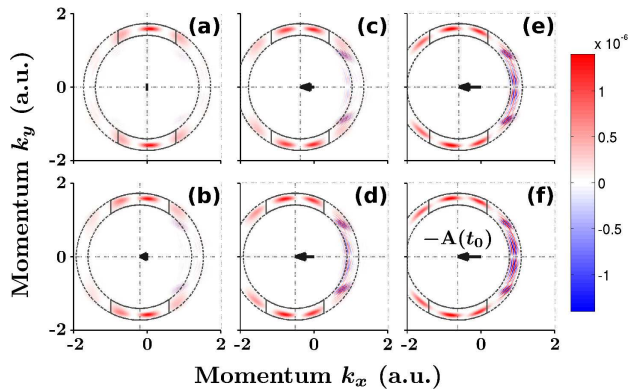}
\caption{XUV photoelectron signals at the XUV-IR delays of
(a) $\Delta t=0.00$
(b) $\Delta t=0.05$
(c) $\Delta t=0.10$
(d) $\Delta t=0.15$
(e) $\Delta t=0.20$
(f) $\Delta t=0.25$ IR cycles. 
($\bfR \parallel \hat{x}$ and $\hat{\bfeps}_{\XUV} \parallel \hat{y}$)
}
\label{fig: Pmom}
\end{center}
\end{figure}

Figure~\ref{fig: Pmom}
shows the
XUV photoionization signal obtained by the numerical simulations
at six delay times from $\Delta t=0$ to $0.25$ IR laser cycles.
As the Gaussian factor in 
the model formula (\ref{eq: photoelectron distribution by cL and cR}) suggests,
the photoelectron momentum is distributed
around the energy-conservation circle of
radius $\sqrt{2[(E_{\text{g}}+E_{\text{u}})/2 + \Omega]}$,
whose center is streaked by the vector potential of the IR laser field as
$\bfk=-\bfA(t_0)$~\cite{prl88_173903}.
This ring distribution is multiplied by the atomic p$_y$-wave factor due to
$|\tilde{d}_{\text{atom}}|^2$, 
as well as the two-center interference pattern from which the ultrafast 
evolution of $c_{\text{L}}(t_0)$ and $c_{\text{R}}(t_0)$ can be retrieved.

\begin{figure}[t]
\begin{center}
\includegraphics[width=.90\linewidth]{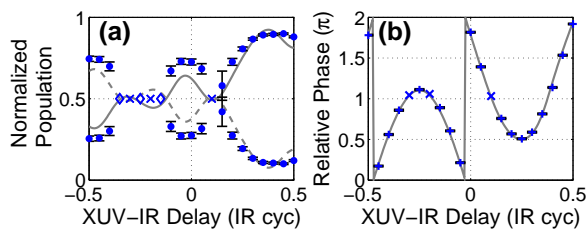}
\caption{
(a) Normalized local populations $P_{\text{L}}$ and $P_{\text{R}}$
and (b) relative phase $\alpha_{\text{LR}}$ retrieved
by the fitting procedure (markers) 
in comparison with the values obtained from the
TDSE solutions (gray lines).
}
\label{fig: quantum state retrieval}
\end{center}
\end{figure}

In actual experiments, it is, however, 
difficult to determine the absolute value of the
photoelectron momentum distribution.
Under this restriction, the information about
the common scale of 
$|c_{\text{L}}(t_0)|$ and $|c_{\text{R}}(t_0)|$ is lost,
but we can still retrieve the normalized populations,
$P_{\text{L}}= |c_{\text{L}}|^2/(|c_{\text{L}}|^2+|c_{\text{R}}|^2)$ and
$P_{\text{R}}= |c_{\text{R}}|^2/(|c_{\text{L}}|^2+|c_{\text{R}}|^2)$,
as well as the relative phase
$\alpha_{\text{LR}} = \arg[c_{\text{L}} c^*_{\text{R}}]$.
Figure~\ref{fig: quantum state retrieval} shows 
these quantities retrieved from the distributions 
in Fig.\ref{fig: Pmom} (and those at other XUV-IR delays). 
We also computed the exact time-evolution of $P_{\text{L}}$, $P_{\text{R}}$,
and $\alpha_{\text{LR}}$ in the
TDSE simulation with only the IR pulse applied, and show the results
in Fig.~\ref{fig: quantum state retrieval} by gray lines as reference.
The comparison shows
that the present method 
allows one to reconstruct the attosecond time-evolution of the laser-dressed
quantum state in H$_2^+$.
We may note that due to the present assumption
that the polarization direction of the XUV
pulse is perpendicular to the molecular axis we obtain
the same momentum distribution when $P_{\text{L}}$ is exchanged with $P_{\text{R}}$.
An extension to other alignment angles
is straightforward, and in this case
the set of values $\{P_{\text{L}}, P_{\text{R}}, \alpha_{\text{LR}}\}$
can be retrieved uniquely.

In summary, we have found that there can be MIBs
from H$_2^+$ and other molecules with CR states 
within a half-cycle of the laser field oscillation
in contrast to the widely accepted tunnel ionization picture. 
These bursts have been shown to be induced by the transient electron localization
inside the molecule on the attosecond time-scale, and 
a simple expression 
to predict the number and instants of the electron localization 
has been presented. 
The time-evolution of the quantum state exhibiting such a localization behavior
can be reconstructed in an experiment with an 
attosecond XUV pulse.

Beyond the paradigm change from 
the tunneling ionization picture, our findings
also suggest that 
in a HHG process, where the intermediate cation is
actually dressed by the intense driving field,
the ultrafast time-evolution of  the electronic state
can be drastically different from that under the 
field-free condition if the cation is in a CR state. This should influence the results of 
techniques to reconstruct molecular 
orbitals~\cite{nature432_867}. 
The MIBs also offer a control knob of the attosecond pulse
generation for which the harmonics at the cut-off of the spectrum is used. 
The photon energy in HHG 
corresponds to 
the kinetic energy of the
electron acquired in the driving field, which
is determined 
by the phase of the field at which the electron is released~\cite{prl71_1994,physRevA49_2117}. 
Our findings suggest that the ionization probability 
at a particular phase, 
and hence the harmonic 
efficiency at the cut-off, can be controlled.

We thank Prof.~H.~Kono, Dr.~F.~He, Dr.~C.~Ruiz M{\'{e}}ndez, and Dr.~T.~Popmintchev
for helpful discussions.
This work was partially supported by NSF.

\bibliography{bib003_takemoto}

\end{document}